\documentclass{article}

\PassOptionsToPackage{numbers, compress}{natbib}

\usepackage[final]{neurips_ai4d3_2023}

\usepackage{tikz}
\usetikzlibrary{shapes.geometric, arrows, chains, positioning}

\usepackage[utf8]{inputenc} 
\usepackage[T1]{fontenc}    
\usepackage{hyperref}       
\usepackage{url}            
\usepackage{booktabs}       
\usepackage{amsfonts}       
\usepackage{nicefrac}       
\usepackage{microtype}      
\usepackage{xcolor}         
\usepackage{graphicx}
\graphicspath{ {./images/} }
\usepackage{subfig}
\usepackage[export]{adjustbox}

\newcommand{\smiles}{\textsc{Smiles}}

\title{Leap: molecular synthesisability scoring with intermediates}

\author{
  Antonia Calvi \\
  Exscientia \\
  \texttt{acalvi@exscientia.co.uk} \\
  \And
  Théophile Gaudin \\
  Exscientia \\
  \texttt{tgaudin@exscientia.co.uk} \\
  \AND
  Dominik Miketa\\
  Exscientia \\
  \texttt{dmiketa@exscientia.co.uk} \\
  \And
  Dominique Sydow \\
  Exscientia \\
  \texttt{dsydow@exscientia.co.uk} \\
  \And
  Liam Wilbraham \\
  Exscientia \\
  \texttt{lwilbraham@exscientia.co.uk} \\
}

\begin{document}

\maketitle

\begin{abstract}
Assessing whether a molecule can be synthesised is a primary task in drug discovery. It enables computational chemists to filter for viable compounds
or bias molecular generative models. The notion of \textit{synthesisability} is dynamic as it evolves depending on the availability of key compounds. A common approach in drug discovery involves exploring the chemical space surrounding synthetically-accessible
intermediates. This strategy improves the synthesisability of the derived molecules due to the availability of key intermediates.
Existing synthesisability scoring methods such as SAScore, SCScore and RAScore, cannot condition on intermediates dynamically.
Our approach, Leap, is a GPT-2 model trained on the depth, or longest linear path, of predicted synthesis routes that allows information on the availability of key intermediates to be included at inference time.
We show that Leap surpasses all other scoring methods by at least 5\% on AUC score when identifying synthesisable molecules, and can successfully adapt predicted scores when presented with a relevant intermediate compound.
\end{abstract}

\section{Introduction}

The ability to predict whether a compound can be made is key to accelerating drug discovery.  Generative methods that neglect synthesisability often suggest compounds that are challenging or impossible to make~\cite{synth-mol-generative}.  Attempting to synthesise these problematic compounds is costly and time-consuming. Therefore, it is crucial to assess synthesisability as part of the scoring functions applied within generative methods to prioritise synthetically accessible compounds.

Furthermore, the concept of synthesisability should be regarded as dynamic and dependent on the compounds available at a specific time. Synthesis planning algorithms~\cite{askos,aizynth,ibmrxn,segler} are used to identify possible synthetic pathways for specific molecules by using readily accessible compounds. However, these methods are computationally expensive and slow to apply exhaustively within generative workflows, which can involve the assessment of hundreds of thousands to millions of compounds~\cite{synth-mol-generative}. Therefore, it makes them unsuitable to assess synthesisability when speed is required to provide feedback to generative models. In practice, this limits the applicability of generative workflows to real world scenarios where an fast, automated and accurate metric of chemical tractability is critical to enable chemists to prioritise high quality compounds that minimise development time and cost. 

To address this challenge, scoring methods such as SAScore~\cite{sascore},
SCScore~\cite{scscore}, and RAScore~\cite{rascore} were developed to assess the synthetic accessibility of molecules.  SAScore assumes that the complexity of synthesising a molecule is correlated with the scarcity of its fragments in databases of bioactive compounds (e.g. PubChem~\cite{pubchem} or ChEMBL~\cite{chembl}) and checks for complex chemistry like stereocenters and ring systems. SCScore is designed to ensure that, on average, the products of chemical reactions exhibit greater synthetic complexity compared to their reactants. RAScore~\cite{rascore} classifies whether a synthetic route can be identified for a particular compound or not by the retrosynthesis tool AiZynthFinder~\cite{aizynth}. During inference, none of these methods are aware of possible available intermediates for the query molecules. In drug discovery projects, a common strategy is to identify, through the synthesis of molecules around a specific chemical space, large quantities of useful synthesisable intermediates from which further compounds can be readily made~\cite{principles-drug-discovery}. When failing to adapt to this information, existing scorers can overestimate the synthetic complexity of molecules produced in this way~\cite{bad-sas}, hindering their ability to identify compounds as synthesisable, particularly when these intermediates are complex.

\begin{figure}[t]
\centering
\includegraphics[width=0.8\textwidth]{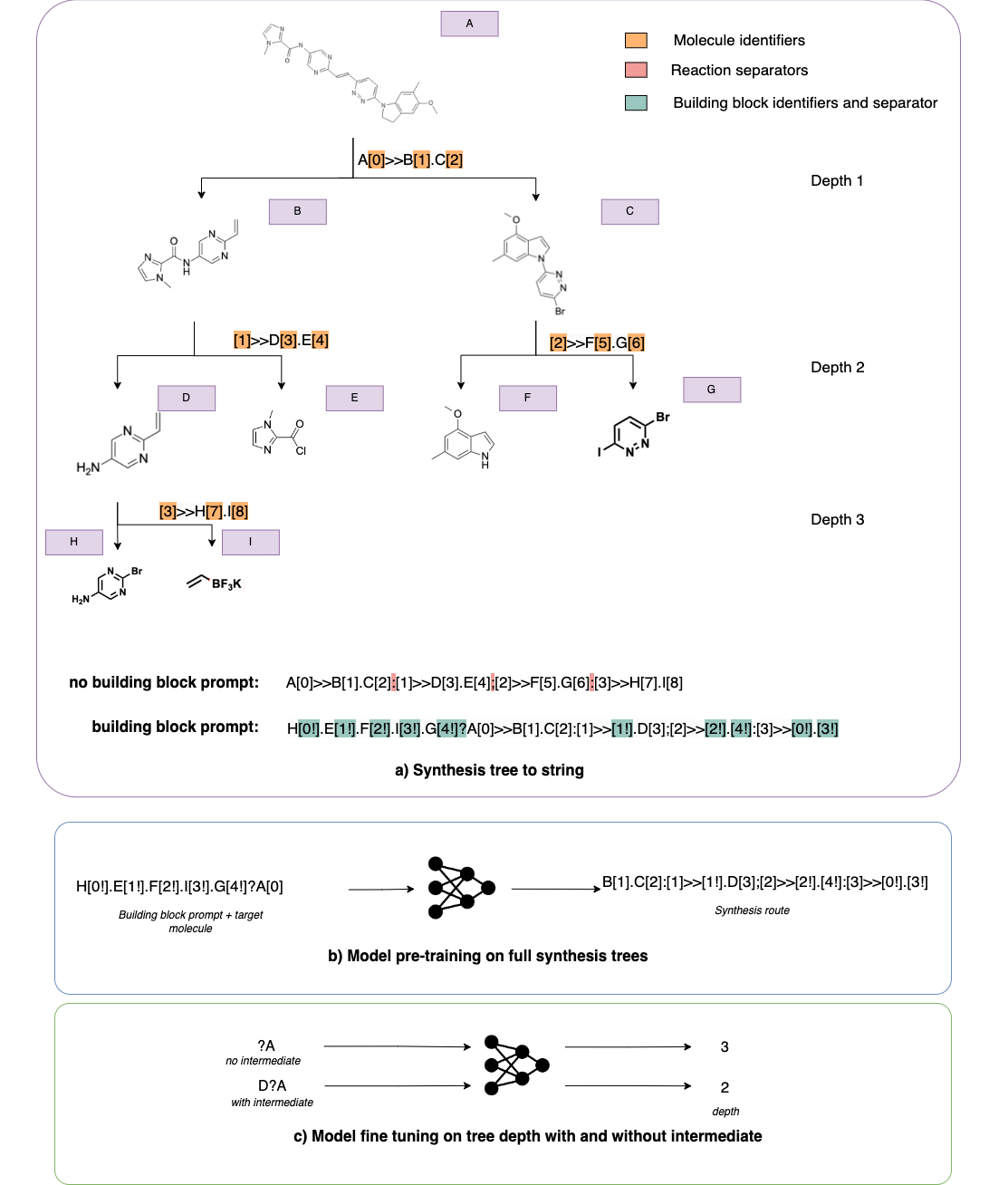}
\caption{Schematic diagram of our workflow. a) shows an example of a synthesis tree of depth 3, and its representation as a string; b) shows input and expected output of our model during pre-training, while c) shows the input and expected output of our model fine-tuned to predict tree depth for a molecule with and without intermediate.}
\label{fig:tree-rep}
\end{figure}



Our main contribution is a novel synthesisability scoring method, Leap, that can adapt to available intermediates to better estimate the practical synthetic complexity of a target molecule. We introduce a specific pre-training objective, where we train a GPT-2 model to predict the entire multi-step retrosynthesis route for a target compound. We use the retrosynthetic planning tool AiZynthFinder to compute the routes of compounds sampled from ChEMBL, and fine-tune the pre-trained model to predict a molecule's synthetic complexity with and without an intermediate. \autoref{fig:tree-rep} shows a schematic representation of our method. 

Two practically useful characteristics distinguish our method from prior art. First, compared to existing scorers, Leap adapts its synthetic complexity score in the presence of useful intermediates for a target molecule. There is currently no existing tool that offers this capability. Second, we demonstrate that even in the absence of intermediates, Leap outperforms existing scorers, providing a strong baseline for future research in this area.

The rest of the paper is organised as follows: in Section \ref{methods} we expand on our method and provide details on both the pre-training and fine-tuning steps; in Section~\ref{experiments} we outline our experiments and discuss the results on both publicly available data and molecules coming from real drug discovery projects; finally we conclude and discuss further work in Section~\ref{conclusion}.

\section{Methods}
\label{methods}

\subsection{Problem Setting}
Within this work, we also refer to a synthesis route as a synthetic tree. \autoref{fig:tree-rep}a. shows an example of this. The root of the tree is the target molecule to be synthesised. The leaves are the building blocks and consist of available compounds. All the other molecules are intermediates. The depth of the tree (tree depth) is given by the maximum number of steps needed to go from the target molecule to any of the building blocks.

Our method is based on the success of transformers~\cite{transformer} in predicting single-step forward reactions (reactants to product), as well as single-step retrosynthesis (product to reactants) ~\cite{molecular-transformer, retroprime} by extending the approach to multi-step routes. We achieve this by encoding synthesis routes as strings, and training a transformer model (GPT-2~\cite{gpt2}) to predict them. To guide the model towards the expected route, we prompt it with the expected building blocks. Finally, we fine-tune this model using a regression head to predict tree depth for target molecules.

\subsection{Pre-Training}

\subsubsection{Data Representation}
\label{data-rep-pre-training}

Our dataset consists of synthesis routes that we represent with a custom string format. This representation encodes both the tree depth and intermediate molecules that act as reactants and products of two reactions as shown in \autoref{fig:tree-rep}b.

Each molecule is encoded as \smiles. Inspired by the reaction smiles~\cite{smirks}, we represent these in a \texttt{product}\texttt{>}\texttt{>}\texttt{reactants} format, where each reactant is separated by a dot.
Reactions occurring at the same depth are separated by a \texttt{;} , while we use \texttt{:} to separate reactions at different depths of the tree. 

Each molecule is given a unique identifier, in the format of \texttt{[0],[1],[2]} etc. These are assigned in ascending order as we traverse the tree from the root to the leaves. The target molecule (i.e. the root of the tree) is always assigned the identifier \texttt{[0]}. When a molecule is encountered for the first time, it is represented by its \smiles ~followed by its identifier. If that same molecule is used in a subsequent reaction, only the identifier is used. This shortens the length of the string, freeing GPT-2's context window from repeated \smiles. \autoref{fig:tree-rep}a. shows an example.

To prompt the model with the expected building blocks for a route, we introduce these at the beginning of the string, separating them from the target molecule with a \texttt{?}. As before, each building block's \smiles ~is followed by a unique identifier. We represent these with \texttt{[0!],[1!],[2!]} etc. Then, when the building blocks are used along the synthesis route, we refer to them using their identifier as shown in \autoref{fig:tree-rep}.

\subsubsection{Model Architecture and Training}
 
Our model is a GPT-2 model with 8 attention heads and 5 hidden layers. We use HuggingFace's implementation~\cite{huggingface}. The hidden states have a dimension of 512, while the feed-forward layers have a dimension of 1024. We train the model for 16 epochs, use AdamW optimiser with a learning rate of 0.0003, and a cosine learning rate scheduler with 2000 warmup steps. We train the model on a language modelling task, where the goal is to predict the next token. 

\subsubsection{Dataset}

To pre-train our model, we use a combination of two datasets. As no dataset of multi-step synthetic routes exists, methods have been developed to generate pathways from existing single-step reaction datasets. The first dataset from ~\citet{metro}, uses dynamic programming to extract synthesis routes from patent reactions found within the USPTO-all~\cite{uspto} dataset. We use train, test and validation splits from \citet{metro} which gives 39717, 5479 and 5493 routes for each.

The dataset from \citet{metro} is skewed both towards short routes of an average depth of three, and routes that only extend down a single branch of the synthesis tree (i.e. linear rather than convergent synthesis routes). To increase diversity, and allow for longer and more branched routes, we extend the training set with a second dataset based on \citet{gao2022amortized}. This models the problem of synthetic tree generation as a Markov decision process with a random policy over reactants and reaction templates. Reaction templates are rule-based patterns extracted from existing reactions to define transformations on molecules that represent valid chemical reactions. From this we obtain 116029 synthesis routes, which we add to the training set. 

We convert all routes to the string format described in Section \ref{data-rep-pre-training}. We augment the data by randomising the \smiles ~as suggested in \citet{shuffle-smiles}, and randomly permuting both the reactions located at the same depth of the route, and the building blocks in the prompt. We apply augmentation three times per route. 

\subsection{Fine-Tuning}

The purpose of fine-tuning the model is to have it predict the expected tree depth of a target molecule both when an intermediate is provided and when it is not. 

\subsubsection{Data Representation}

The input to the model consists of the target molecule and, when available, an intermediate. Both molecules are represented with their respective \smiles. 
The target molecule is prepended by the \texttt{?} token that is used to separate it from the potential intermediate. Therefore, the input format to the model will either be \texttt{intermediate?target} or \texttt{?target} depending on whether an intermediate is included. 

The output of the model is an integer that corresponds to the expected tree depth. A simplified example of this is shown in \autoref{fig:tree-rep}c.

\subsubsection{Model Architecture and Training}

We add a regression head on top of the pre-trained GPT-2 model to predict tree depth. We tune the hyperparameters by hand.  Fine-tuning only the regression head, and the the regression head with last two hidden layers, showed worse performance on the validation set than fine-tuning all layers of the model. We use an AdamW optimizer with a learning rate of 0.0001, and a linear scheduler with 4000 warmup steps. We minimise the mean squared error, and train for 16 epochs with early stopping on the validation loss. 

\subsubsection{Dataset}
\label{data-fine-tuning}

The fine-tuning dataset is a sample of 290418 ChEMBL~\cite{chembl} molecules that we use to train our model. In addition, we curate a held-out dataset of 4989 project molecules coming directly from internal drug discovery projects, to challenge the model on out-of-domain prediction. 

We use AiZynthFinder~\cite{aizynth} with default parameters to compute the routes of both datasets. A route is considered to be solved if all leaves are purchasable molecules. If more than one route is found for a molecule, the route scored the highest by AiZynthFinder is retained. For solved routes, we compute the depth of the tree. Since the maximum tree depth allowed by AiZynthFinder's default parameters is 7, we assign a depth of 10 for molecules where a route is not found. 

To create target molecule-intermediate pairs, we randomly sample a maximum of three intermediate molecules for each route and recompute the depth accordingly. We do the latter by treating the sampled intermediate as an available building block, meaning that we can ignore the synthesis steps needed to create it. At a synthesis tree level, this effectively results in the removal of any nodes beyond the intermediate molecule. This reduces the depth of the tree when the intermediate is found along the longest branch of the tree.

To train the model to recognise that not every molecule is a viable intermediate and pathologically reduce predicted route depth in response to supplied intermediates, we also generate a set of "false" intermediates. We do this by pairing target molecules with a random molecule sampled across all intermediates in the dataset. For these negative pairs, the expected depth is the same as that of the target molecule when no intermediate is supplied.

ChEMBL molecules are randomly split 70/15/15 into train, test, validation sets, leaving 202421, 43563, 44434 molecules, respectively. After augmentation we have a total of 646421, 139594, 142600 data points per set. The project molecules are augmented in the same way, totalling 20277 project-intermediate data points.

\section{Experiments}
\label{experiments}

We run experiments to assess the ability of Leap to adapt to the knowledge of intermediates, its ability to classify the synthetic feasibility of molecules, and whether pre-training helps. We compare to SAScore, SCScore and RAScore. SAScore is normalised between 0 and 1 through min-max scaling. We explore results on the ChEMBL test set as well as internal molecules set to assess generalisation to potentially out-of-domain novel compounds.

\subsection{Identifying synthesisable molecules}

\begin{figure}[t]
\centering
\subfloat[Test molecule]{
  \includegraphics[width=0.5\textwidth]{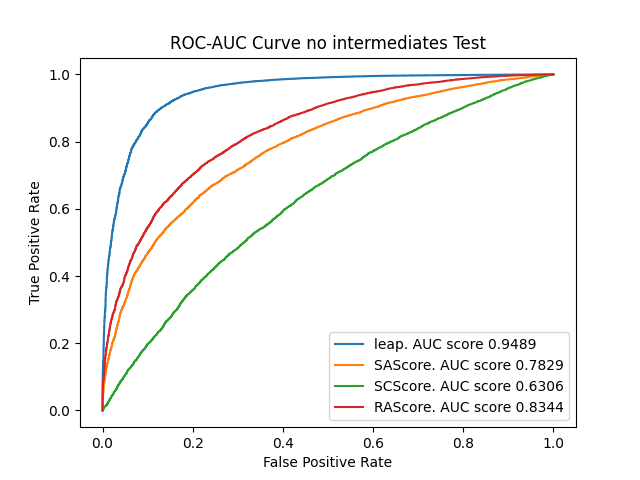}
}
\subfloat[Project molecules]{
  \includegraphics[width=0.5\textwidth]{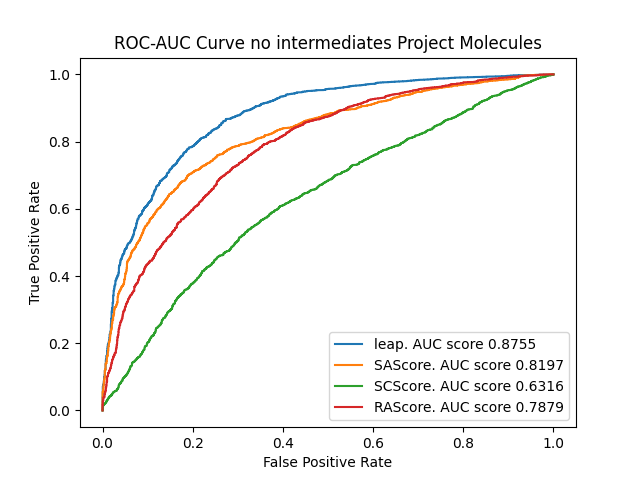}
}
\caption{ROC-AUC curves for all scorers on both test and project molecules.}
\label{fig:roc-auc}
\end{figure}

\begin{figure}[htp]
\centering
 \includegraphics[width=\textwidth]{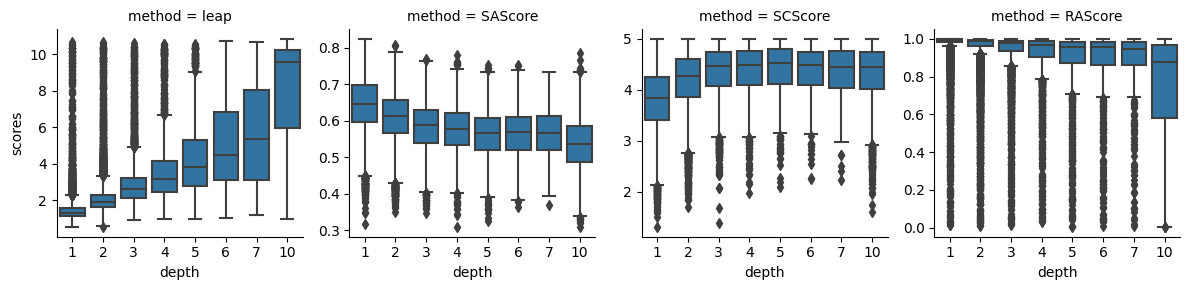}
\caption{Distribution of scores assigned by Leap, SAScore, SCScore and RAScore for molecules with synthetic routes of different depths.}
\label{fig:test_boxplot}
\bigskip
\includegraphics[width=\textwidth]{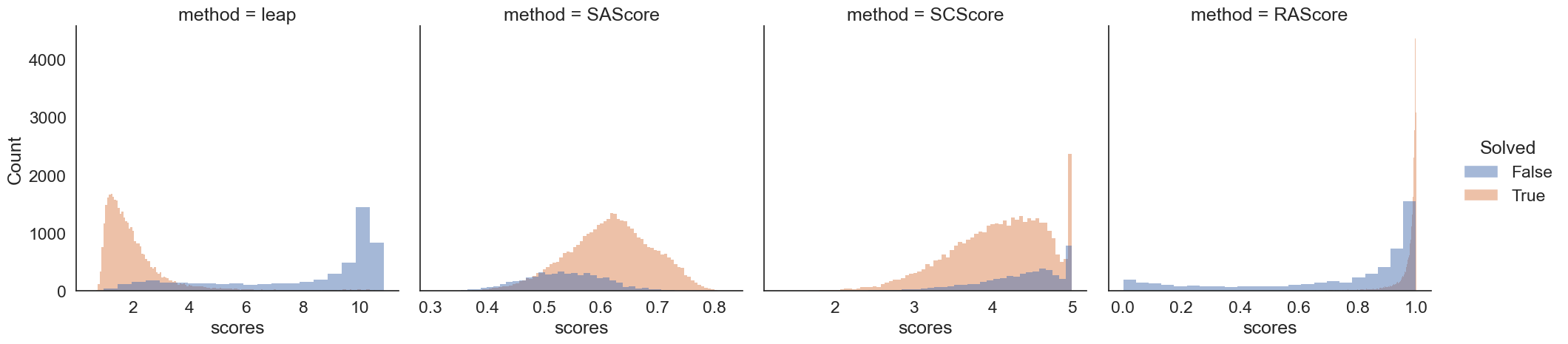}
\caption{Distribution of scores assigned by Leap, SAScore, SCScore and RAScore for solved and unsolved compounds.}
\label{fig:distribution_scores}
\end{figure}

To determine molecular synthesisability we compute routes using AiZynthFinder~\cite{aizynth} for both test and project molecules. If a route can be found, the molecule is classified as synthesisable.

To permit comparison between Leap and RAScore, SAScore and SCScore, and to establish a baseline for further experiments, we do not consider intermediates in this experiment. We score all molecules with all scorers, and use these to compute the ROC-AUC curves shown in \autoref{fig:roc-auc}. Leap outperforms all other methods in terms of AUC score for both test and project molecules. This demonstrates that our model remains useful even when intermediates are not present. Leap also shows superior performance on the hold-out set of project molecules, showing its ability to generalise beyond its training domain. Conversely, even though RAScore was also trained on ChEMBL molecules, its predictive performance deteriorates on out-of-domain molecules as previously observed in \citet{rascore}.

We hypothesise that the ability of our model to better distinguish between easy- and difficult-to-synthesise molecules is due to Leap using tree depth as a proxy for synthetic complexity. We show, from Figure \ref{fig:test_boxplot} that scores predicted by Leap correlate with the tree depth predicted by AiZynthFinder. All other scorers show very similar scores across molecules that require different depths of synthesis. Therefore, this can make it harder to establish the optimal threshold to differentiate between synthesisable and non-synthesisable molecules: as shown in \autoref{fig:distribution_scores} Leap is the only scorer that shows a clear distinction in scores assigned to synthesisable and non-synthesisable molecules.

\subsection{Effects of intermediates}

\begin{figure}[h]
\centering
\includegraphics[width=10cm]{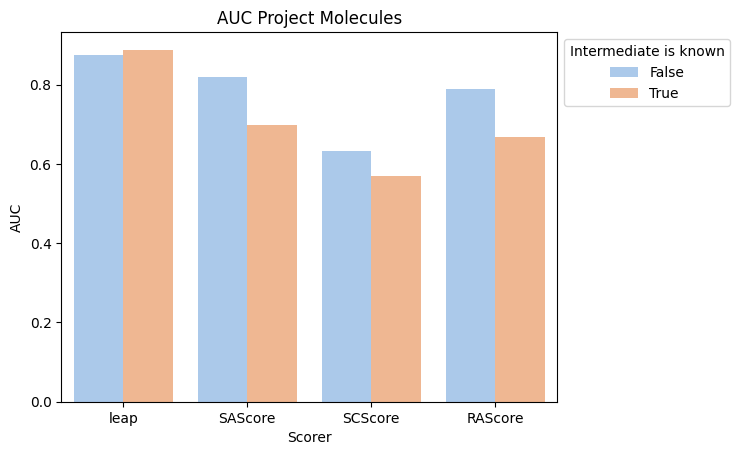}
\caption{Barplot showing the AUC score for the various scorers when we do and do not have a key intermediate for molecules.}
\label{fig:barplots-auc-intermediate}
\end{figure}

In the previous experiments, we showed that even when not supplying intermediates Leap performs better than other scorers in identifying whether a molecule is synthesisable. Here, we explore how Leap performs when intermediates are supplied. We use the test and project molecule sets obtained from Section \ref{data-fine-tuning}, both with and without intermediates. We then score all the molecules with the scorers, however only Leap can make use of intermediates. From Section \ref{data-fine-tuning}, we mark molecules as synthesisable if their assigned depth is not 10 and non-synthesisable otherwise. Note that the same molecule can go from non-synthesisable to synthesisable if a useful intermediate is supplied. It should also be noted that not all intermediates have this effect: aside from including false intermediates, selecting intermediates at random can result in some of them not contributing to the reduction of tree depth.

\begin{figure}[t]
\centering
\subfloat[Test molecule]{
  \includegraphics[width=0.5\textwidth]{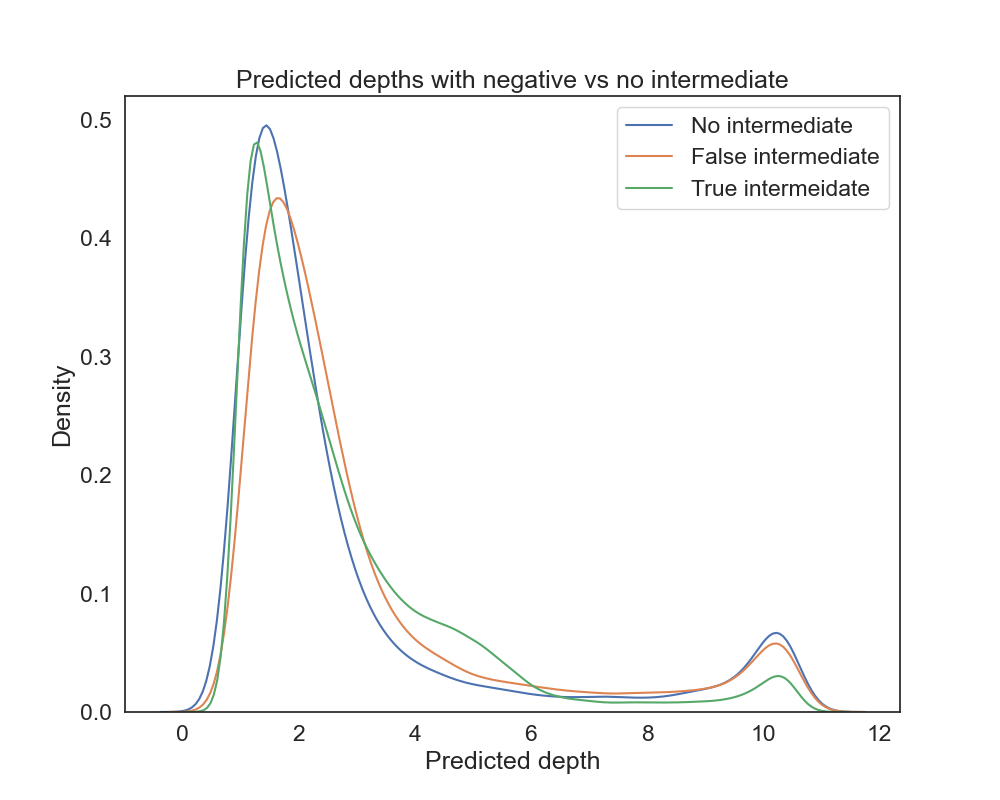}
}
\subfloat[Project Molecules]{
  \includegraphics[width=0.5\textwidth]{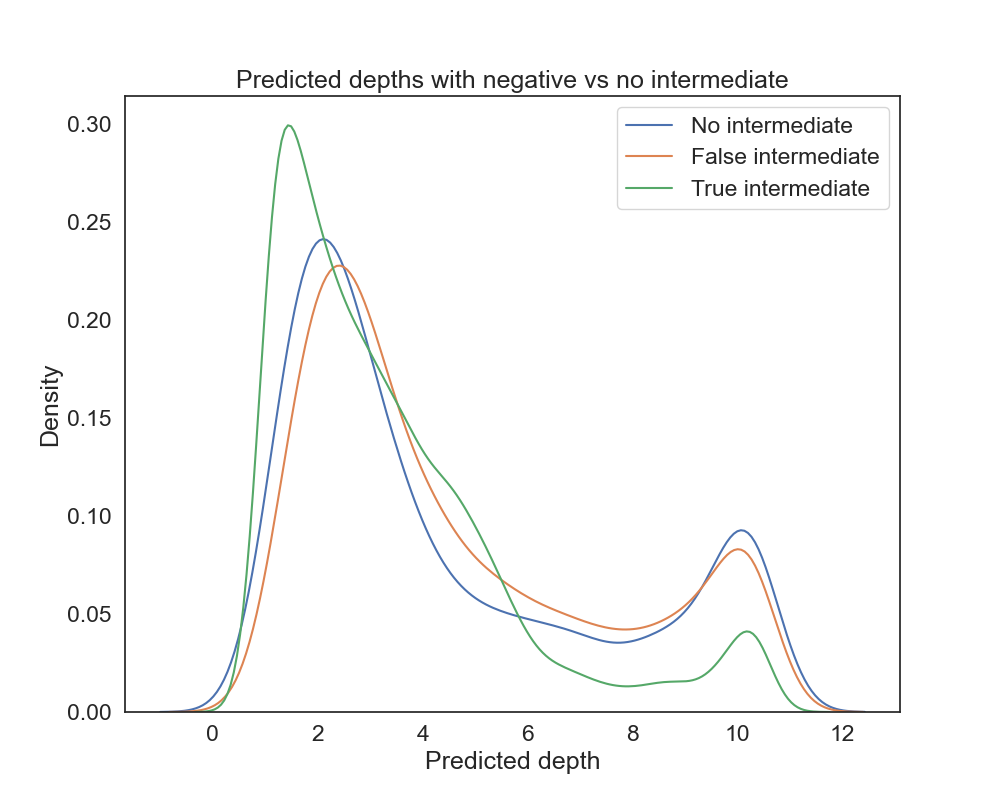}
}
\caption{Distributions of predicted depths for both project and test molecules when false, true and no intermediates are supplied.}
\label{fig:neg-intermediates}
\end{figure}

We compute the AUC for each scorer in identifying whether a molecule is synthesisable, both when no intermediate is known, as well as when an intermediate is provided. From \autoref{fig:barplots-auc-intermediate} we can see that for molecules for which a key intermediate is given, Leap can differentiate between synthesisable molecules with an AUC score of 0.89. Conversely, all other scorers incur a decrease in their AUC scores. This is to be expected, but highlights the limitations of such scorers: complex synthesis routes become simpler if chemists already possess key intermediates. We further check that our model does not pathologically lower the predicted score if any intermediate is supplied by comparing the distributions between supplying false and no intermediates. \autoref{fig:neg-intermediates} shows that distributions of predicted scores are very close for both test and project molecules, indicating that Leap's predictions are robust against irrelevant intermediates.

\subsection{Ranking molecules with intermediates}

The purpose of this experiment is to investigate whether our model maintains the ranking of the expected tree depth as different intermediates are supplied. We group both test and project sets by their target molecules, and then compare the expected ranking of tree depth for each target molecule with the different intermediates, to that obtained by the prediction of our model. If everything is ranked correctly, then that is considered a correct ranking. For this experiment, we do not consider false intermediates. We obtained 77\% correct ranking on the test set, and 69\% on the project set.

\subsection{Comparison to baseline models}

\begin{table}[t]
\centering
\begin{tabular}{lllllll}
\toprule
                                         & \multicolumn{3}{c}{Test}                                           & \multicolumn{3}{c}{Project}                   \\ 
\multicolumn{1}{c|}{Model}               & R2            & MAE           & \multicolumn{1}{l|}{MSE}           & R2            & MAE           & MSE           \\ \midrule
\multicolumn{1}{l|}{CatBoost}            & 0.42          & 1.32          & \multicolumn{1}{l|}{3.63}          & 0.19          & 1.98          & 7.18          \\
\multicolumn{1}{l|}{BERT}                & 0.54          & 1.05          & \multicolumn{1}{l|}{2.89}          & 0.29          & 1.78          & 6.77          \\
\multicolumn{1}{l|}{GPT-2} & 0.41          & 1.23          & \multicolumn{1}{l|}{3.71}          & 0.17          & 1.92          & 7.48          \\
\multicolumn{1}{l|}{Leap}                & \textbf{0.64} & \textbf{0.88} & \multicolumn{1}{l|}{\textbf{2.29}} & \textbf{0.41} & \textbf{1.56} & \textbf{5.59} \\
\bottomrule
\end{tabular}
\caption{Comparison between performance of our model Leap to the baseline models on regression metrics: R2, mean absolute error (MAE), mean squared error (MSE)}
\label{tab:baselines}
\end{table}

To validate our architecture and our pre-training method, we conduct experiments using three baseline models: CatBoost, GPT-2 (i.e. Leap with no pre-training) and BERT~\cite{bert}. We compare their performance on R2, mean absolute error (MAE) and mean squared error (MSE). For CatBoost, intermediate and target molecules are converted to ECFP4 fingerprints, and then concatenated. For BERT, they are passed in as two sentences, separated by \texttt{SEP} token. The results are reported in Table \ref{tab:baselines}. We show that pre-training helped boost the performance of Leap, and that the fine-tuned Leap outperforms both CatBoost and BERT on all the metrics.

\section{Conclusion}
\label{conclusion}

We introduced Leap, the first method for scoring the synthesisability of molecules that considers intermediates. We have pre-trained a model on the task of generating synthesis routes, and fine-tuned it on tree depth prediction with and without intermediates. We have shown the ability of our model to generalise to out-of-domain molecules and to take advantage of intermediates provided from various points along a synthesis route. Future work includes testing this scorer paired with a generative model, as well as testing the effect of more challenging false intermediates, for example by modifying intermediates along a synthesis route of a target molecule.

\medskip

\bibliographystyle{plainnat}
\bibliography{neurips_ai4d3_2023}


\end{document}